\documentclass[aps,12pt,superscriptaddress,amsfonts,amssymb,amsmath]{revtex4}

\usepackage{graphicx}
\usepackage{epsfig}
\usepackage{makeidx}
\usepackage{epstopdf}

\begin{document}

\title{Design of a test for the electromagnetic \\ coupling of non-local wavefunctions}

\author{G.\ Modanese \footnote{Email address: giovanni.modanese@unibz.it}}
\affiliation{Free University of Bolzano-Bozen \\ Faculty of Science and Technology \\ I-39100 Bolzano, Italy}

\linespread{0.9}

\begin{abstract}

\bigskip

It has recently been proven that certain effective wavefunctions in fractional quantum mechanics and condensed matter do not have a locally conserved current; as a consequence, their coupling to the electromagnetic field leads to extended Maxwell equations, featuring non-local, formally simple additional source terms. Solving these equations in general form or finding analytical approximations is a formidable task, but numerical solutions can be obtained by performing some bulky double-retarded integrals. We focus on concrete experimental situations which may allow to detect an anomalous quasi-static magnetic field generated by these (collective) wavefunctions in cuprate superconductors. We compute the spatial dependence of the field and its amplitude as a function of microscopic parameters including the fraction $\eta$ of supercurrent that is not locally conserved in Josephson junctions between grains, the thickness $a$ of the junctions and the size $\varepsilon$ of their current sinks and sources. The results show that the anomalous field is actually detectable
at the macroscopic level with sensitive experiments, and can be important at the microscopic level because of virtual charge effects typical of the extended Maxwell equations.

\end{abstract}

\maketitle

\section{Introduction}

In some recent works \cite{Modanese2017MPLB,modanese2017electromagnetic,hively2012toward,van2001generalisation,jimenez2011cosmological,arbab2017extended} an extension of Maxwell equations has been proposed, which makes them applicable also to systems where charge is conserved globally but not locally. This extension is based on an idea originally expressed by Aharonov and Bohm; as shown in \cite{Modanese2017MPLB}, it is the only possible relativistically invariant extension of the standard electromagnetic massless Lagrangian and leads to covariant modified Maxwell equations which are formally simple and appealing. When the equations are written in the usual 3D vector formalism one immediately recognizes in the equations for $\nabla\cdot \textbf{E}$ and $\nabla\times\textbf{B}$, besides the usual terms, two additional terms that are retarded integrals of the ``extra-source'' $I=\partial_t \rho+\nabla\cdot\textbf{J}$. This quantity is of course vanishing in the usual approach, where one supposes the validity of the continuity equation $\partial_t \rho +\nabla\cdot\textbf{J}=0$.

As pointed out in \cite{modanese2017electromagnetic,modanese2018time}, sources with wavefunctions $\Psi$ that do not satisfy a continuity equation are generally possible in quantum mechanics. Such sources are not present in the standard formalism based on the Schr\"odinger equation (and its nonlinear extensions) or in quantum field theory with local interactions. Nevertheless, non-locally conserved currents arise for some effective wavefunctions, like those describing nuclear scattering \cite{chamon1997nonlocal,balantekin1998green}, systems with long range interactions and anomalous diffusion \cite{Lenzi2008solutions,latora1999superdiffusion,caspi2000enhanced}, superconductors in the general non-local Gorkov theory \cite{waldram1996superconductivity,hook1973ginzburg}, or fractional quantum mechanics \cite{wei2016comment,Lenzi2008fractional}.

The electromagnetic coupling of such wavefunctions requires an extension of the usual Maxwell theory. The extension satisfies a general principle of ``censorship'' \cite{Modanese2017MPLB}, according to which the electromagnetic field generated by any non-conserved source is still equivalent to the field of some suitable conserved source; the difference between the real source and the fictitious conserved source is that the latter is not limited in space, so in certain cases the non-conservation is effectively censored, but in others it is not, and there can be physical consequences. The censorship principle can be seen as a safeguard of the strict locality of the electromagnetic field, even when it is coupled to quantum systems with ``spooky'' non-locality (extending Einstein's famous judgment to wavefunctions with non-local equations).

In this work we compute numerical solutions of the extended equations in the presence of a specific anomalous source, chosen in view of a possible experimental verification of the new theory (Sects.\ \ref{ret}, \ref{num}). Namely, we consider a brief pulse of supercurrent which crosses multiple normal barriers thanks to the proximity effect, in the assumption that the macroscopic wavefunction of the superconducting charge carriers satisfies a Ginzburg-Landau equation with non-local terms \cite{hook1973ginzburg}. We find the anomalous corrections to the magnetic field of the supercurrent and propose a scheme for a possible detection procedure (Sect.\ \ref{exp-test}). We make recourse to numerical solutions because the mentioned extra-source terms in the extended Maxwell equations, though formally simple, are very difficult to evaluate, except for static cases with high symmetry, as done in \cite{Modanese2017MPLB}. This is also because some standard approximations like the multipole expansion do not apply in this case, although other approximations can probably be found with more advanced mathematical approaches \cite{fabrizio2003electromagnetism}. Sect.\ \ref{con} contains our conclusions.

\section{Retarded integrals for the field of a current pulse in a junction}
\label{ret}

Let us first recall the extended Maxwell equations, in CGS units. All field and sources are functions of $(t,\textbf{x})$, also if not explicitly denoted. Define the extra-source $I(t,\textbf{x})$ as the function which quantifies the violation of local current conservation:
\begin{align}
	I(t,\textbf{x})=\frac{\partial\rho}{\partial t}+\nabla\cdot\textbf{J}
\end{align}

In the familiar 3D vector formalism the extended equations without sources are written as usual, namely $\nabla \times \textbf{E}=-(1/c)(\partial \textbf{B}/\partial t)$, $\nabla \cdot \textbf{B}=0$. The extended equations with sources take the form
\begin{align}
	\nabla \cdot \textbf{E}=4\pi \rho-\frac{1}{c^2}\frac{\partial}{\partial t}\int d^3y \frac{I\left(t_{ret},\textbf{y} \right)}{\left|\textbf{x}-\textbf{y} \right|}; \label{eqE}
\end{align}
\begin{align}
	\nabla \times \textbf{B}-\frac{1}{c} \frac{\partial \textbf{E}}{\partial t}=\frac{4\pi}{c} \textbf{J}+\frac{1}{c} \nabla \int d^3y \frac{I\left(t_{ret},\textbf{y} \right)}{\left|\textbf{x}-\textbf{y} \right|}. \label{eqB}
\end{align}
where $I_{ret}=I(t-|\textbf{x}-\textbf{y}|/c,\textbf{x})$.

The solution of these (linear) equations can be written in the form
\begin{align}
	\textbf{E}=\textbf{E}^0+\textbf{E}^s, \qquad \textbf{B}=\textbf{B}^0+\textbf{B}^s
\end{align}
where $\textbf{E}^0$, $\textbf{B}^0$ are the solutions with $I=0$. We call $\textbf{E}^s$ and $\textbf{B}^s$ the anomalous electric and magnetic contributions.

Although the extended theory is not gauge invariant \cite{Modanese2017MPLB,jimenez2011cosmological}, at the mathematical level the equations (\ref{eqE}), (\ref{eqB}) can still be solved by introducing auxiliary potentials (thanks to the equations for $\nabla \times \textbf{E}$ and $\nabla \cdot \textbf{B}$ and to uniqueness theorems \cite{woodside2009three}). We denote by $\phi_{aux}$ and $\textbf{A}_{aux}$ auxiliary potentials in the Feynman-Lorenz gauge. They satisfy the equations
\begin{align}
	\frac{1}{c^2}\frac{\partial^2 \phi_{aux}}{\partial t^2}-\nabla^2 \phi_{aux}=4\pi \rho-\frac{1}{c^2} \frac{\partial}{\partial t} \int d^3y \frac{I\left(t_{ret},\textbf{y} \right)}{\left|\textbf{x}-\textbf{y} \right|}; \label{pot1} \end{align}
\begin{align}
	\frac{1}{c^2}\frac{\partial^2 \textbf{A}_{aux}}{\partial t^2}-\nabla^2 \textbf{A}_{aux}=\frac{4\pi}{c} \textbf{J}+\frac{1}{c} \nabla \int d^3y \frac{I\left(t_{ret},\textbf{y} \right)}{\left|\textbf{x}-\textbf{y} \right|}. \label{pot2}
\end{align}
The relation between potentials and fields is as usual: $\textbf{E}=-c^{-1}\partial_t \textbf{A}_{aux}-\nabla \phi_{aux}$, $\textbf{B}=\nabla \times \textbf{A}_{aux}$.  

From eqs.\ (\ref{pot1}), (\ref{pot2}) we can write expressions for the anomalous electric and magnetic contributions $\textbf{E}^s$ and $\textbf{B}^s$. First we solve eqs.\ (\ref{pot1}), (\ref{pot2}) by writing $\phi^s_{aux}$ and $\textbf{A}^s_{aux}$ (the auxiliary potentials generated by the extra-current $I$) as double-retarded integrals.
We set $k=c^{-1}$:
\begin{align}
	\phi^s=\frac{1}{4 \pi} \int \frac{d^3y}{|\textbf{x}-\textbf{y}|} \left[ -k^2 \frac{\partial}{\partial t} \int \frac{d^3z}{|\textbf{y}-\textbf{z}|} I\left(t - k|\textbf{y}-\textbf{z}|,\textbf{z} \right) \right]_{t \to t-k|\textbf{x}-\textbf{y}|}
	\label{phiI}
\end{align}
\begin{align}
	\textbf{A}^s=\frac{1}{4 \pi} \int \frac{d^3y}{|\textbf{x}-\textbf{y}|} \left[ k \nabla_y \int \frac{d^3z}{|\textbf{y}-\textbf{z}|} I\left(t - k|\textbf{y}-\textbf{z}|,\textbf{z} \right) \right]_{t \to t-k|\textbf{x}-\textbf{y}|}
	\label{AI}
\end{align}

In order to compute $\textbf{B}^s$ we take the curl of $\textbf{A}^s$. The curl operator can be eventually brought under the integral in $d^3y$, but before this one has to perform, as written in (\ref{phiI}), (\ref{AI}), (a) the first retardation (expressed in the argument of $I$), (b) the gradient in $\textbf{y}$, (c) the second retardation (denoted by the subscript of the square bracket). Similarly, to compute $\textbf{E}^s$ we take $-k\partial_t \textbf{A}^s-\nabla \phi^s$, after the same three steps.

As discussed in the Introduction, we do not see at present any general approximation scheme apt to make these integrals tractable analytically. The results of the numerical evaluations described below actually indicate that the presence of an extended ``cloud'' of secondary charge and current makes it difficult to use multipole expansions or similar techniques. Therefore we shall focus our attention on a concrete example of source $I$, which may allow experimental measurements of the anomalous magnetic field $\textbf{B}^s$, and we shall evaluate the fields numerically.

\begin{figure}
\begin{center}
\includegraphics[width=11cm,height=8cm]{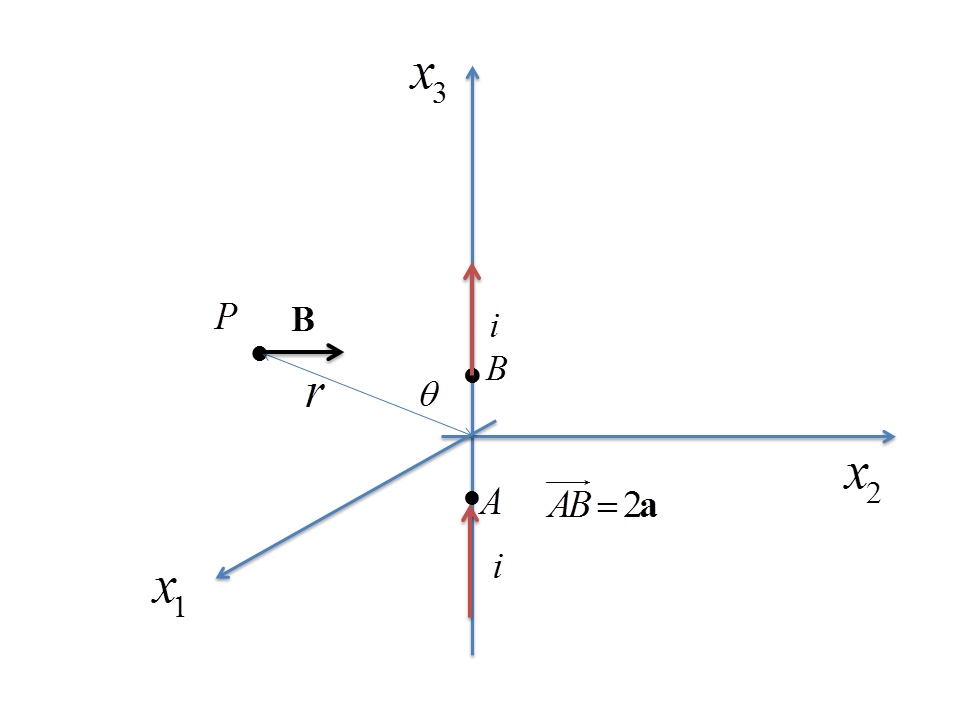}
\caption{Geometrical configuration for the calculation of the anomalous magnetic field. A current $i$ flows in a conductor with negligible section along $x_3$ and is interrupted between $A$ and $B$, with violation of the continuity condition at $A$ and $B$. The component $B_2^s$ of the anomalous field generated by this interruption is computed at the point $P$, placed on the plane $x_1$-$x_3$, with distance $r$ from the origin and azimuthal angle $\theta$. The ``missing regular field'' $B_2^0$ is given by the Biot-Savart formula $B_2^0=2ia\sin\theta/(cr^2)$.} 
\label{fig-assi}
\end{center}  
\end{figure}

\begin{figure}
\begin{center}
\includegraphics[width=12cm,height=8cm]{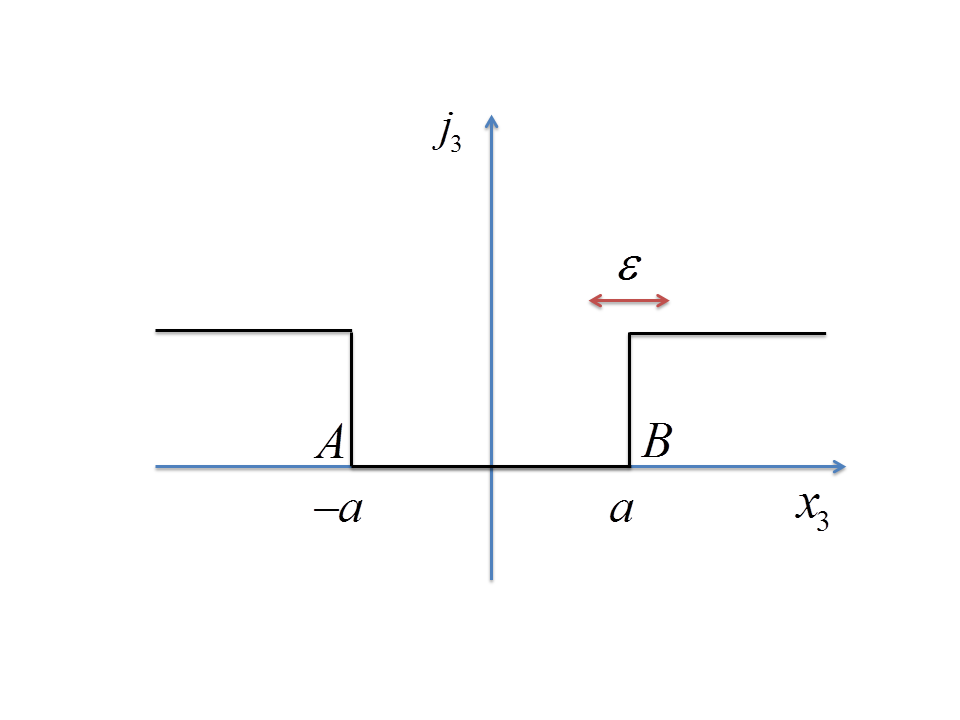}
\caption{Behavior of the ``strong tunnelling'' current density $j_3$ as a function of $x_3$ in the case of an abrupt interruption. In the actual calculation the step at $\pm a$ is regularized by smoothing it out with a cutoff of magnitude order $\varepsilon$. Therefore $\varepsilon$ is the effective size of the charge sink/source located respectively at $A$ and $B$.} 
\label{fig-corrente}
\end{center}  
\end{figure}

Consider a straight wire carrying a current pulse with duration of order $\tau$ (Fig.\ \ref{fig-assi}). Suppose that the wire passes through the origin $O$ of a coordinate system, and that there is a tunnelling barrier for the charge carriers centered at $O$. Let $i$ be the total current. Suppose that a certain part $i_s$ of the total current $i$ ($i_s \ll i$) crosses the barrier by anomalous, ``strong'' tunnelling: this means that this current violates the local conservation condition and there is a point $A$ at the wire in position $-\textbf{a}$ where the current $i_s$ vanishes, re-appearing at $B$, in position $+\textbf{a}$. (This is a simplified formal representation of a wavefunction with local non-conservation, compare \cite{modanese2018time}.) Let us also admit that $\rho=0$ everywhere, so that we have only current and no net charge. We choose a reference system where the current flows along $z$, so $\textbf{a}=(0,0,a)$. Taking for instance a Gaussian form for the time dependence, the only non-zero component of the current density is written as
\begin{align}
	j^s_z=i_s e^{-\frac{t^2}{2 \tau^2}} \delta(x)\delta(y) \left[ \theta(-a-z) + \theta(z-a) \right]
\label{eq9}
\end{align}
The dependence of this density on $z$ (apart from the factor $\delta(x)\delta(y)$) is depicted in Fig.\ \ref{fig-corrente}.

The extra-source $I$ is
\begin{align}
	I=\frac{\partial\rho}{\partial t}- \nabla \cdot \textbf{j}=i_s e^{-\frac{t}{2\tau^2}}\left[ \delta^3 (\textbf{x}-\textbf{a})-\delta^3 (\textbf{x}+\textbf{a}) \right]
	\label{eqI}
\end{align}

In the following we shall consider a pulse duration $\tau$ of the order of $10^{-5}$ to $10^{-3}$ s; this allows to disregard not only the charge density $\rho$, but also $\partial_t \rho$, because even if the tunnelling junction has a small stray capacitance, its effect at these frequencies is very small.

\section{Numerical evaluation}
\label{num}

\subsection{Method. Finite source size}
\label{met}

Our aim is to compare the anomalous field $\textbf{B}^s$ generated by the extra-source (\ref{eqI}) with the regular field $\textbf{B}^0$ of the current ($B^0=2ia\sin\theta/(cr^2)$). Like $\textbf{B}^0$, $\textbf{B}^s$ has cylindrical symmetry. Let us compute it in the plane $x_1$-$x_3$ (Fig.\ \ref{fig-assi}). We look for the component $B_2^s$, to be compared with $B_2^0$. We start from the vector potential (\ref{AI}). First of all, let us compute the retarded integral in $d^3z$, with the source (\ref{eqI}). We obtain
\begin{align}
	\int \frac{d^3z}{|\textbf{y}-\textbf{z}|} I\left(t_{ret},\textbf{z} \right) =
	i_s \left[ \frac{e^{-\frac{1}{2\tau^2} \left( t-\frac{1}{c}|\textbf{y}-\textbf{a}| \right)^2}}{|\textbf{y}-\textbf{a}|} - \frac{e^{-\frac{1}{2\tau^2} \left( t-\frac{1}{c}|\textbf{y}+\textbf{a}| \right)^2}}{|\textbf{y}+\textbf{a}|}
	\right] \label{1ret}
\end{align}

Next we compute the derivatives of this expression with respect to $y_1$ and $y_3$, in order to obtain the components $A_1$ and $A_3$ of the vector potential. The resulting expressions are retarded in time and multiplied by $1/|\textbf{x}-\textbf{y}|$. Then we differentiate the first expression with respect to $x_3$, the second with respect to $x_1$, and take the difference, in order to find the second component of the curl of $\textbf{A}$. Note that the lengths of vectors like $(\textbf{y}\pm\textbf{a})$ must be expressed in terms of the components, for example
\begin{align}
	|\textbf{y}-\textbf{a}|=\sqrt{(y_1-a_1)^2+(y_2-a_2)^2+(y_3-a_3)^2}
\end{align}
The explicit algebraic expressions obtained in this way (and further including a cut-off for the size of the source as specified below) are very long and are handled with \texttt{Mathematica}. Finally we replace the parameters $a$ and $\tau$ with their numerical values (see below for the choice of $a$), replace the coordinates $x_1$, $x_2$, $x_3$ according to the configuration in Fig.\ \ref{fig-assi}, choose a value for $t$ (typically $t=\tau=10^{-5}$) and perform numerically the integral in $d^3y$. The values of the distance $r$ and angle $\theta$ at which the field is computed are varied (see results in Sect.\ \ref{results}).

\begin{figure}
\begin{center}
\includegraphics[width=12cm,height=9cm]{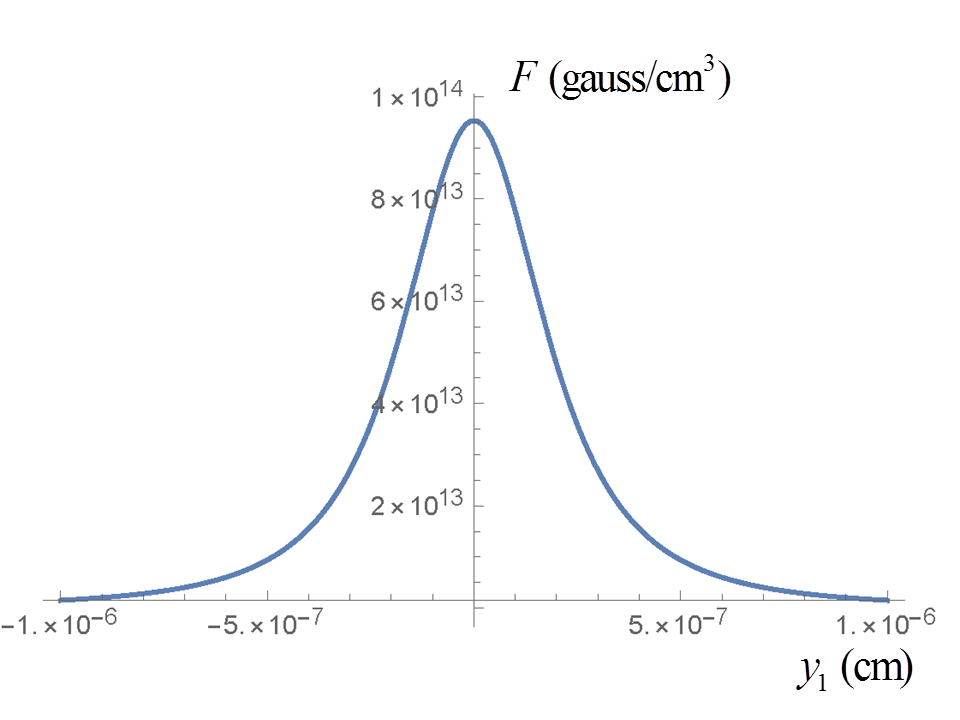}
\caption{Dependence on $y_1$ of the integrand $F$ of the double-retarded integral which gives the anomalous field ${\bf B}^s$ as curl of ${\bf A}^s$ in eq.\ (\ref{AI}). The complete expression of $F$ is given in the Appendix. The coordinates $y_2$ and $y_3$ are set to zero in this graph. One notices that the effective field source is enlarged with respect to the primary current flowing along the $x_3$ axis and whose density has a $\delta$-function in $y_1$. The curve shown here decreases as $y_1^{-3}$. Such a ``cloud'' of secondary current is typical of the extended Maxwell equations, in the presence of sources which violate the local continuity condition.} 
\label{fig1}
\end{center}  
\end{figure}

\begin{figure}
\begin{center}
\includegraphics[width=12cm,height=9cm]{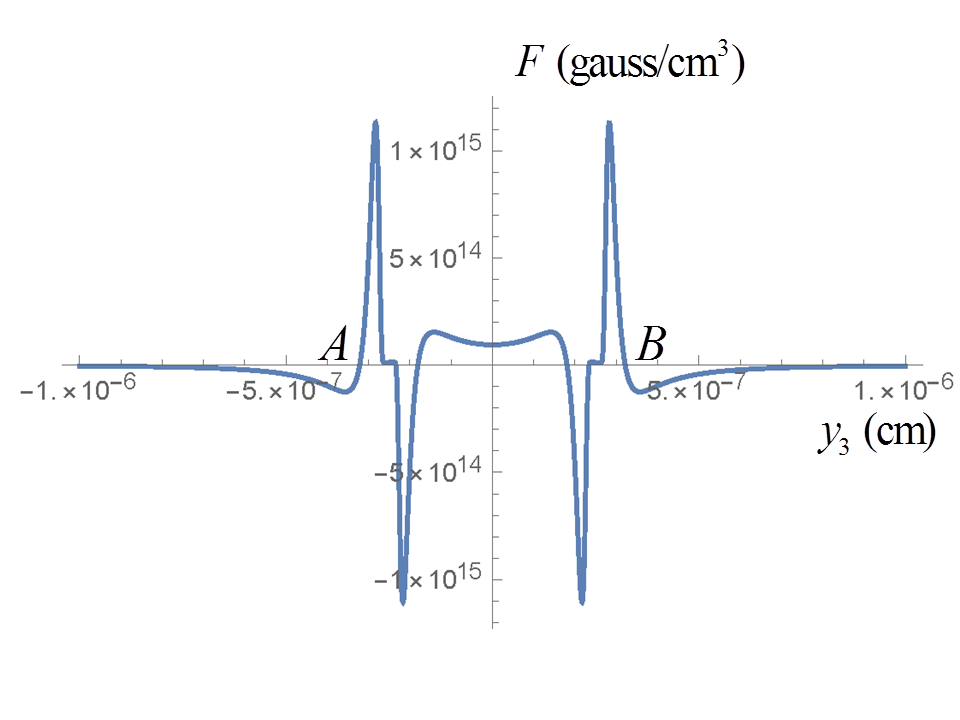}
\caption{Dependence on $y_3$ of the integrand $F$ of the double-retarded integral which gives the anomalous field ${\bf B}^s$ as curl of ${\bf A}^s$ in eq.\ (\ref{AI}). The complete expression of $F$ is given in the Appendix. The coordinates $y_1$ and $y_2$ are set to zero in this graph. Each pair of positive and negative peaks corresponds to one of the $\delta$-functions in the extra-source $I$ which violates local conservation, namely A corresponds to the current sink and B to the current source. The $\delta$-functions have been regularized with a finite width $\varepsilon$, which is also the width of the peaks. The distance between A and B is equal to $2a$ (``length of strong tunnelling'').
} 
\label{fig3}
\end{center}  
\end{figure}

The integrals contain functions with sharp peaks due to the localized sources (see below for the corresponding cutoffs) and long power-law  tails due to the secondary current (see Figs.\ \ref{fig1}, \ref{fig3}). It is therefore necessary to split the integration region into several domains, in order to obtain a reliable result with the function \texttt{NIntegrate} of \texttt{Mathematica}. The \texttt{Working Precision} parameter is gradually increased until the results are stable, typically to the value 14 or 15. The integration range at infinity has also been gradually extended until the results stabilize. As a check, we also have  computed the integrals with a Monte Carlo algorithm; in its code the definition of the integrand can be subdivided into several parts, so the code looks cumbersome but is compact and complete, and we have reported it in the Appendix. (In contrast, the expanded integrand used by \texttt{Mathematica} is too long to be reported.) Each Monte Carlo integration requires typically one day, while each numerical integration with \texttt{Mathematica} requires a few minutes.

Finally, some specifications are in order concerning a cut-off that is needed to account for the finite size of the source $I$. The $\delta$-function in (\ref{eqI}) is formally convenient and allows to perform analytically the first retarded integration. This integration, however, gives us a formal analogue of the electric potential of a dipolar source, and therefore contains non-integrable singularities for $\textbf{y} \to \pm \textbf{a}$. In order to regularize them, we introduce a finite size $\varepsilon$ of the source, i.e., physically, of the region where $\partial_t\rho +\nabla\cdot\textbf{J} \neq 0$. Therefore $\varepsilon$ is the size of the ``sink'' where the current $i_s$ of strong tunnelling disappears; this size is to be compared with the tunnelling length $a$, i.e., the distance between the points of disappearance and re-appearance of the current). It is known that the potential $V(r)$ generated by a charged spherically symmetric body of radius $\varepsilon$ reaches its maximum at $r=\varepsilon$ and then decreases to zero when $r\to 0$. So we enforce a smooth cut-off on the retarded integral of $I$ in (\ref{1ret}) by multiplying the first term by $\exp(-\varepsilon^2/4|\textbf{y}-\textbf{a}|^2)$ and the second term by $\exp(-\varepsilon^2/4|\textbf{y}+\textbf{a}|^2)$. The choice of $\varepsilon$ does not significantly affect the results, as explained below.

\subsection{Results}
\label{results}

The aim of the numerical computation described in Sect.\ \ref{met} was to find the ratio $B^s/B^0$ as a function of the distance $r$ and azimuthal angle $\theta$ (see Fig.\ \ref{fig-assi}). The field $B^s$ is the anomalous field generated by the extra-source (\ref{eqI}), while $B^0$ is the normal Biot-Savart field $B^0=(2ai/c)\sin\theta/r^2$ that would be generated by the current element of length $2a$ if it would flow normally, respecting the continuity equation. The microscopic parameters $a$, $\varepsilon$ were chosen as $a=2.5\cdot 10^{-7}$ cm, $\varepsilon=1.0\cdot 10^{-7}$ cm, as motivated in Sect.\ \ref{exp-test}.

Quite surprisingly, the ratio $B^s/B^0$ was found to vanish for any $\theta$, in all the explored range of $r$ (from $10^{-6}$ to $10^{-1}$ cm), due to a cancellation of contributions from different integration regions. This cancellation occurs at different distances, depending on $r$ (see Tab.\ \ref{tab1}). It can be interpreted as due to the combined effect of the extra-source localized at the sink/well $A$ and $B$, and of the secondary current density which forms a ``cloud'' in space decreasing as $|\textbf{y}|^{-3}$ (compare Figs.\ \ref{fig1}, \ref{fig3}). The cancellation occurs independently from the values of $a$ and $\varepsilon$.

The ratios $B^s/B^0$ given in Tab.\ \ref{tab1} are computed assuming the same generating current for $B^s$ and $B^0$. In fact, it is reasonable to expect that only a small fraction $\eta$ of the current of the junction crosses the junction by strong tunnelling, i.e., without local conservation. For example, the non-local corrections to the Ginzburg-Landau equation in the proximity effect are small, of the order of 1\% or less of the order parameter \cite{hook1973ginzburg}. As a consequence, the effect of the missing $B^s$ field on the total field will be small, of the same magnitude order.

Moreover, the volume of the junctions is a small fraction of the the total volume of the material. For instance, in a sintered YBCO sample with average grain size of the order of 10 $\mu$m and inter-grain junctions of average thickness 0.1 $\mu$m, the volume of the junctions is of the order of $10^{-2}$ of the total volume. The total field generated will be in general a complicated integral on the whole material, but one could predict in the example above an overall variation of the order of 1 part in $10^4$, with respect to a normal material (a material without any strong tunnelling and local non-conservation).

At the macroscopic level the effect of field reduction due to the strong tunnelling is therefore hard to detect, but we shall nevertheless propose in the next Section a possible measurement method, based on three parallel wires with equal and opposite currents, one of which may host strong tunnelling close to $I_c$, so that the field measured at the midpoints between the wires is not exactly zero.

\begin{table}[h]
\begin{center}
\begin{tabular}{|c|c|c|c|c|}
\hline
\textbf{Integr.\ region} & $r=10^{-6}$ cm & $r=10^{-5}$ & $r=10^{-4}$ & $r=10^{-3}$  \\
\hline
$[0,0.5]\cdot 10^{-6}$ cm & 0.180 & 0.208 & 0.208 & 0.208  \\
$[0.5,5]\cdot 10^{-6}$ & -0.179 & -0.023 & -0.006 & -0.006  \\
$[5,25]\cdot 10^{-6}$  & -0.001 & -0.178 & -0.004 & -1$\cdot 10^{-4}$ \\
$[25,125]\cdot 10^{-6}$  & -8$\cdot 10^{-6}$ & -0.008 & -0.127 & -0.001  \\
$[125,625]\cdot 10^{-6}$  & & -6$\cdot 10^{-5}$ & -0.071 & -0.023  \\
$[0.625,3]\cdot 10^{-3}$  & &  & -5$\cdot 10^{-4}$ & -0.174  \\
$[3,15]\cdot 10^{-3}$  & &  &  & -0.005  \\
$[15,75]\cdot 10^{-3}$  & &  &  & -3$\cdot 10^{-5}$  \\
\hline
$B^s/B^0$ & 0.00 & 0.00 & 0.00 & 0.00 \\
\hline
\end{tabular}
\end{center}
\caption
{Contributions to the adimensional ratio $B^s(r)/B^0(r)$ between anomalous field and Biot-Savart field, for four values of $r$. The contributions arise from eight different integration regions of the retarded integral (\ref{AI}) (the appropriate curl derivatives need also to be inserted into the integral; see details in the text). For each value of $r$, the sum of the contributions gives zero within errors. The contribution of the first region, which contains the physical source (from 0 to $0.5\cdot 10^{-6}$ cm) is always positive, but is canceled by the integration on the cloud of secondary current, over regions which are located further away from the physical source as $r$ grows. The results for $r=10^{-2}$ cm and $r=10^{-1}$ cm, not shown here, are completely analogous. Each integration region, except for the first one, is subdivided into 26 sub-regions, namely the 3-intervals obtained when $y_1$, $y_2$ and $y_3$ vary between the values given in the table. The microscopic parameters are fixed to $a=2.5\cdot 10^{-7}$ cm and $\varepsilon=1.0\cdot 10^{-7}$ cm. The ratio $B^s(r)/B^0(r)$ does not depend on the angle $\theta$. Data obtained with the \texttt{NIntegrate} function of \texttt{Mathematica} and checked with sample Monte Carlo integrations.}
\label{tab1}
\end{table}

\section{Experimental test}
\label{exp-test}

The numerical evaluations of the previous Section could be applied, in a superconductor, to a narrow weak link or a Josephson junction where the largest part $i^0$ of the tunnelling supercurrent obeys the continuity equation and generates a regular field $B^0$, but a small fraction $i^s$ of the supercurrent does not obey the continuity equation and generates an anomalous magnetic field $B^s$ (which is practically zero, at the frequencies corresponding to our pulse, compared to $B^0$).

In order to test this theoretical model, detecting the missing field or possibly setting an upper limit on the quantity $\eta=i^s/i^0$, we think more specifically of a high-$T_c$ superconducting material like YBCO carrying a supercurrent. It is well known that in such materials the current flows across a large number of intrinsic and inter-grain junctions \cite{kleiner1994intrinsic,hilgenkamp2002grain}. It is also known that tunnelling in these junctions cannot be described by the BCS theory and one has to resort to a phenomenological Ginzburg-Landau theory or more exactly to the non-local Gorkov theory for the proximity effect \cite{wang2001continuous,waldram1996superconductivity}. The macroscopic coherence of superconducting wavefunctions may play a crucial role in amplifying nonlocal effects, which in other cases, like for fractional quantum mechanics \cite{modanese2018time}, remain confined at a microscopic level.

The microscopic parameters of the tunnelling process depend strongly on the details of the material and its preparation \cite{hilgenkamp2002grain}. We can suppose, for instance, that if the grains have size $l \simeq 1 \mu$m, the inter-grain junctions have thickness $2a \simeq 5 - 10$ nm and the size $\varepsilon$ of the source/sink regions where local conservation fails is of the order of the coherence length $\xi \simeq 1$ nm.

\begin{figure}
\begin{center}
\includegraphics[width=11cm,height=8cm]{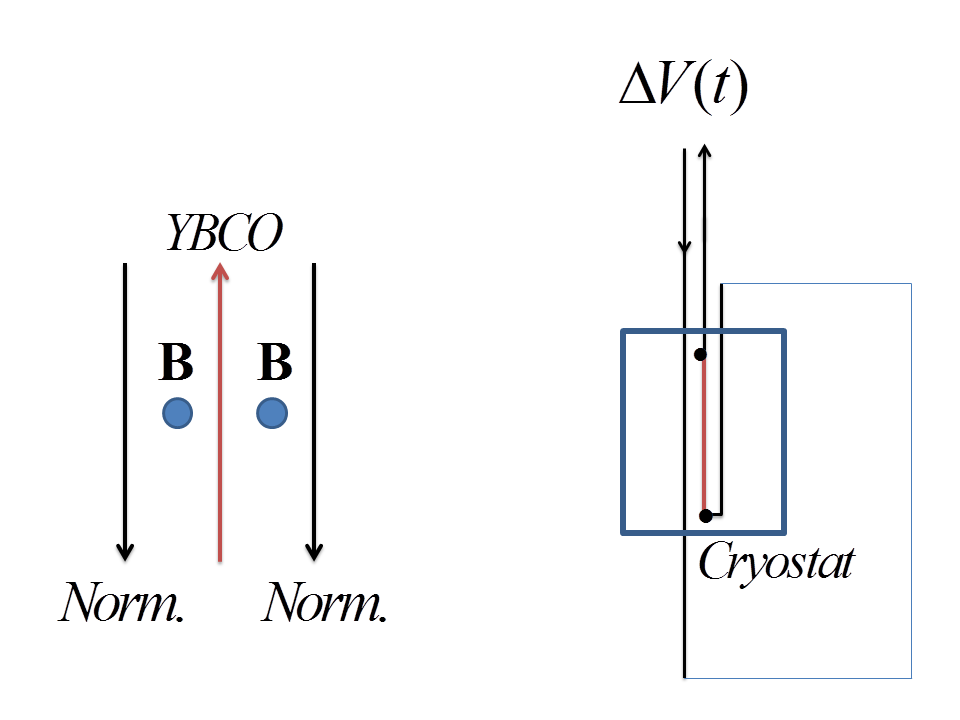}
\caption{Proposed method for the detection of small transient differences between the magnetic field of an YBCO wire (middle, red) and the fields of two normal wires with the same current. Two field probes (${\bf B}$) are placed equidistant from the wires, possibly allowing further fine adjustment of their position. The symmetric configuration prevents lateral displacements of the supercurrent flow in the central wire. See the main text for details of the circuit which generates the brief current pulse. The distance between the wires is of the order of $\sim 1$ cm, their diameter $\sim 1$ mm, the max current $\sim 1$ A. The black dots denote normal-superconducting junctions. 
} 
\label{wires}
\end{center}  
\end{figure}

Of course, small anomalies in the field strength generated by the current in a conductor can have more mundane causes, in particular near $T_c$ or near $I_c$, where the onset of dissipation makes it difficult to keep the current constant. We therefore propose a differential measuring device with three parallel wires with equal and opposite current (Fig.\ \ref{wires}), where the external wires are made of normal metal, and the central one of sintherized YBCO with diameter $\simeq 1$ mm. The distance of the wires should be no more than a few centimeters, also in order to reduce the size of the cooling system.
 
The reason for using three wires, instead of two, is the following: when the YBCO wire is in the superconducting state, its internal current pattern may vary; in particular, with only two wires we may expect that the repulsive magnetic force exerted by the normal wire causes a slight deformation of the current density, such that the pairs density decreases near the side of the wires which faces the other wire. This deformation would depend in general on the temperature and total current and therefore it might interfere with the effect we want to observe. For a magnitude order estimate, consider the field $B \propto 1/r$ generated by the YBCO wire on the detector, where $r$ is the distance between the center of the current flow in YBCO and the detector itself. If the center of the current is shifted by, say, 0.1 mm (in a wire of diameter 1 mm) due to the magnetic repulsion, then the relative variation of $B$ will be $\Delta B/B \simeq \Delta (r^{-1}) \cdot \Delta r / r^{-1} \simeq \Delta r/r$. For $r \simeq 10$ cm, this gives $\Delta B/B \simeq 10^{-4}$, which is of the same order of the expected anomaly. A distance $r$ greater than 10 cm would be unpractical for cooling, field measurement etc. Actually, $r \simeq 1$ cm is more realistic. So one must reduce this possible ``asymmetry'' effect by balancing the magnetic repulsion with another normal wire. This also has the advantage of allowing a cross check between two simultaneous magnetic field measurements on opposite sides of the YBCO wire.

The critical current will depend on the details of the material, but let us assume it to be of the order of 1 A \cite{cha1998critical}. As we shall explain shortly, we want to approach $I_c$ and $T_c$ in the measurements. The magnetic field should be measured between the two wires, either with a pick-up coil or a Hall probe. There are two options:

\begin{enumerate}
\item Suppose the position of the probes can be finely adjusted, until the measured field is zero (within errors), when the current is far from the critical current, and at a temperature far from the critical temperature. Then the current or temperature are changed, approaching $I_c$ or $T_c$. A tiny net field should be observed at this moment, if the YBCO wire hosts a small fraction of anomalous current, because of the missing field effect discussed above.

\item Alternatively, the probes can be fixed as precisely as possible in the middle, so that the field residual will be small. According to Maxwell equations, this residual must be exactly proportional to the current, if there is no anomaly in the field generation in the YBCO wire. If this is not the case (near $I_c$ or $T_c$), one should observe deviations from the proportionality between field and current.
\end{enumerate}

There are many possible reasons for why the ratio $\eta=i^s/i^0$, and the magnetic anomalies, could depend on the ratio $I/I_c$ or $T/T_c$. The critical current $I^J_c$ of an inter-grain junction (which is related to $I_c$ but not necessarily equal) depends on its thickness $d$ as $I^J_c \propto e^{-d/\xi}$ \cite{waldram1996superconductivity}. This may be seen as the consequence of a continuity condition $\rho v =const.$, because the pairs density decreases exponentially across the junction and their upper velocity is limited. As $I$ approaches $I_c$, this mechanism is stretched to its limit. The same happens when $T$ approaches $T_c$, and the coherence length $\xi$ diverges. From this intuitive reasoning it is hard to conclude whether the ratio $\eta=i^s/i^0$ should be larger close to $I_c$ and $T_c$ or far from them, but it appears in any case that there may be a dependence, which should show up in the differential measurements.

Note that near $I_c$ YBCO usually begins to show dissipation effects due to internal flux flow \cite{kunchur1995novel,waldram1996superconductivity}.
The flux-flow resistive state of a superconductor just above $I_c$ is a transient state, and this also motivates our choice of short current pulses. In the numerical estimates of the previous section we took the pulse duration as $\tau=10^{-5}$ s. It is possible to check that the computations hold, without any essential modification, also for longer times, for instance $\tau \simeq 10^{-4} - 10^{-3}$ s, but for the measurements such times are probably too long, with the risk of overheating and material damaging. We will therefore stick to a pulse duration $\tau=10^{-5}$. 

Let us briefly design the external circuit which should generate the pulse. We suppose for simplicity to have an RLC circuit near critical damping. The peak discharge current is $I_{peak} \simeq V_0/R$, where $V_0$ is the charging voltage of the capacitor. In principle it would be possible to obtain $I_{peak} \simeq 1$ A as required by keeping $V_0$ very low, provided the load resistance $R_L$ of the circuit is very small: for instance, with $R_L \simeq 10^{-3} \ \Omega$, one has $V_0 \simeq 10^{-3}$ V. (We use MKS units here.) But we should also take into account the contact resistance between the supply cables and the YBCO wire. 

Let us then suppose, more realistically, that the charging voltage is of the order of 1 V and the load resistance of the order of 1 $\Omega$. (This also gives a better impedance matching; the maximum transferred power is of the order of 1 W.) The discharge time $\tau$ depends on the total inductance $L$ as $\tau=L/R$. For $\tau \simeq 10^{-5}$ s, one needs $L \simeq 10^{-5}$ H. The capacitance $C$ needed for the discharge is of the order of $L/R^2$, i.e., $C \simeq 10^{-5}$ F. The discharge circuit is therefore simple to mount and can be switched electronically.

\section{Conclusions}
\label{con}

This work explores the possible consequences, at the macroscopic electromagnetic level, of a failure of the local current conservation in fractional and non-local quantum mechanics. For this purpose, we have applied the extended Aharonov-Bohm electrodynamics to a junction where a small fraction of the current is supposed to flow by ``strong tunnelling''. The mathematical treatment is robust, because our relativistic formulation of the Aharonov-Bohm electrodynamics is uniquely defined and the solution of the equations through a double-retarded integral does not require any special approximation.

The numerical integration described in Sect.\ \ref{num} shows that the secondary current induced by the strong tunnelling of the primary current in the junction does not generate any anomalous field $B^s$. The normal Biot-Savart field $B^0$ of such a primary current is also obviously missing. This is an important result, and the resulting ``missing field'' effect may be observable, even if small compared to the field of the bulk material. Mathematically, this result depends on the assumption that the primary current flow has negligible transversal size (junction seen as a thin wire). In the opposite geometrical limit (strong tunnelling between infinite planes), we have shown instead in \cite{Modanese2017MPLB} that the anomalous field of a stationary current completely replaces the missing Biot-Savart field. In practice, since any real junction has a non-negligible transversal size, we expect that a non-zero anomalous field will be present, but it will amount only to a fraction of the missing Biot-Savart field. Further computations in progress confirm this expectation. In any case, the magnitude order of the missing field does not change, being still of the order of $10^{-4}$ of the bulk field under the microscopic assumptions of Sect.\ \ref{results} (strong tunnelling current $\sim$ 1\% of the total current and junctions volume $\sim$ 1\% of the bulk volume). 

A field anomaly of this magnitude order (or even smaller, say of 1 part in $10^6$) would have already been detected, if it would occur in stationary conditions and with normal materials. However, the conditions for strong tunnelling with local non-conservation are likely to be present only in transient form, and near the critical current or critical temperature in junctions like for instance the inter-grain junctions in YBCO. In such conditions fields vary due to many other causes. For this reason we have proposed a differential measurement of the magnetic fields generated by YBCO and normal wires carrying the same current, as described in Sect.\ \ref{exp-test}. Due to its symmetry, the proposed setup allows to spot any deviation from the integrated fourth Maxwell equation (field circuitation depends only on surface integral of the current), be it of the form predicted by the extended Maxwell equations, or even of a more general form.

\bigskip

\noindent
{\bf Acknowledgment} - This work was supported by the Open Access Publishing Fund of the Free University of Bozen-Bolzano.

\section{Appendix}

We report here the C code used for the Monte Carlo integration. This is useful because it contains the complete expression of the retarded integrals, which are far too long to write in explicit analytical form or as \texttt{Mathematica} output. The macroscopic parameters $r$ and $\theta$ used in this code are $r=0.3$ cm, $\theta=\pi/2$. The cutoff \texttt{epsq4} corresponds to the quantity $\varepsilon^2/4$ mentioned in Sect.\ \ref{met}. The cutoff \texttt{eps1} is a cautionary cutoff on the quantity $|\textbf{x}-\textbf{y}|$ and is actually irrelevant since the integral is convergent when $|\textbf{x}-\textbf{y}|\to 0$. Note that the integrand $F$ below (function of $\textbf{y}$) must be further multiplied by $k/(4\pi)$. Plots of $F$ in dependence on $y_1$ and $y_3$ are given in Figs.\ \ref{fig1}, \ref{fig3}.

\medskip

\textbf{Parameters:}

\texttt{t = 1.0e-5;
a = 2.5e-7;
a1 = 0.0; a2 = 0.0; a3 = a;
epsq4 = pow(5.0e-8,2);
eps1 = 1.0e-7;
tau = 1.0e-5;
g = -1/(2.0*pow(tau,2));
c = 3.0e10;
k = 1.0/c;
x1 = 0.3*sin(Pi/2.0); x2 = 0.0; x3 = 0.3*cos(Pi/2.0);}

\medskip 

\textbf{Sub-functions of the integrand:}

\medskip 

\texttt{X=pow(-a1 + y1,2) + pow(-a2 + y2,2) + pow(-a3 + y3,2);
Y=pow(a1 + y1,2) + pow(a2 + y2,2) + pow(a3 + y3,2);
Z=pow(eps1,2)+pow(x1 - y1,2) + pow(x2 - y2,2) + pow(x3 - y3,2);
SX=sqrt(X);
SY=sqrt(Y);
SZ=sqrt(Z);
X2=pow(X,2);
Y2=pow(Y,2);
X15=pow(SX,3);
Y15=pow(SY,3);
Z15=pow(SZ,3);
EX=exp(-(epsq4/X) + g*pow(t - k*SZ - k*SX,2));
EY=exp(-(epsq4/Y) + g*pow(t - k*SZ - k*SY,2));
}

\medskip

\textbf{Integrand}:

\medskip

\texttt{F = -(((x3 - y3)*(-((EX *(-a1 
+ y1))/X15) + (EY* (a1 + 
y1))/Y15 + (EX* ((2*epsq4*(-a1 + y1))/X2 - 
 (2*g*k*(-a1 + y1)*(t - k*SZ -  k*SX))/SX))/ SX - 
 (EY* ((2*epsq4*(a1 + y1))/Y2 - 
 (2*g*k*(a1 + y1)*(t - k*SZ -  k*SY))/SY))/ SY))/Z15) + 
 ((2*EX*g* pow(k,2)*(-a1 + y1)*(x3 - y3))/ 
 (SZ*X) - (2*EY*g* pow(k,2)*(a1 + y1)*(x3 - 
y3))/(SZ*Y) + (2*EX*g*k* (-a1 + y1)*(x3 - y3)*(t - k*SZ - 
 k*SX))/ (SZ*X15) - (2*EY*g*k* (a1 + y1)*(x3 
- y3)*(t - k*SZ - k*SY)) /(SZ*Y15) - 
 (2*EX*g*k* (x3 - y3)*(t - k*SZ - k*SX)* 
 ((2*epsq4*(-a1 + y1))/X2 - (2*g*k*(-a1 + 
y1)*(t - k*SZ - k*SX))/SX))/ (SZ*SX) + 
 (2*EY*g*k* (x3 - y3)*(t - k*SZ - k*SY)* 
 ((2*epsq4*(a1 + y1))/Y2 - (2*g*k*(a1 + y1)*(t 
- k*SZ - k*SY))/ SY))/ (SZ*SY))/ SZ + 
 ((x1 - y1)*(-((EX* (-a3 + y3))/X15) + 
 (EY* (a3 + y3))/Y15 + (EX* ((2*epsq4*(-a3 
+ y3))/X2 - (2*g*k*(-a3 + y3)*(t - k*SZ - 
 k*SX))/SX))/ SX - (EY* ((2*epsq4*(a3 + y3))/Y2 - 
 (2*g*k*(a3 + y3)*(t - k*SZ -  k*SY))/SY))/ SY))/Z15 - 
 ((2*EX*g* pow(k,2)*(x1 - y1)*(-a3 + y3))/ 
 (SZ*X) - (2*EY*g* pow(k,2)*(x1 - y1)*(a3 + 
y3))/(SZ*Y) + (2*EX*g*k* (x1 - y1)*(-a3 + y3)*(t - k*SZ - 
 k*SX))/ (SZ*X15) - (2*EY*g*k* (x1 - y1)*(a3 
+ y3)*(t - k*SZ - k*SY)) /(SZ*Y15) - 
 (2*EX*g*k* (x1 - y1)*(t - k*SZ - k*SX)* 
 ((2*epsq4*(-a3 + y3))/X2 - (2*g*k*(-a3 + 
y3)*(t - k*SZ - k*SX))/SX))/ (SZ*SX) + 
 (2*EY*g*k* (x1 - y1)*(t - k*SZ - k*SY)* 
 ((2*epsq4*(a3 + y3))/Y2 - (2*g*k*(a3 + y3)*(t 
- k*SZ - k*SY))/ SY))/ (SZ*SY))/ SZ;
}

\bibliographystyle{unsrt}
\bibliography{mme}

\end{document}